\documentstyle[prl,aps,epsf]{revtex}

\begin{document}
\draft
\title { Elliptic Flow: Transition from out-of-plane to in-plane
	Emission in Au~+~Au Collisions }
\author{
C.Pinkenburg$^{(1)}$, N.~N.~Ajitanand$^{(1)}$, J.~M.~Alexander$^{(1)}$,
M.~Anderson$^{(5)}$, D.~Best$^{(*2)}$,
F.P.~Brady$^{(5)}$, T.~Case$^{(2)}$, W.~Caskey$^{(5)}$, D.~Cebra$^{(5)}$,
J.L.~Chance$^{(5)}$, P.~Chung$^{(1)}$, B.~Cole$^{(12)}$, K.~Crowe$^{(2)}$,
A.~C.~Das$^{(3)}$, J.E.~Draper$^{(5)}$, A.~Elmaani$^{(1)}$,
M.L.~Gilkes$^{(1)}$, S.~Gushue$^{(1,9)}$, M.~Heffner$^{(5)}$,
A.S.~Hirsch$^{(7)}$, E.L.~Hjort$^{(7)}$, L.~Huo$^{(14)}$, M.~Justice$^{(4)}$,
M.~Kaplan$^{(8)}$, D.~Keane$^{(4)}$, J.C.~Kintner$^{(13)}$, J.~Klay$^{(5)}$,
D.~Krofcheck$^{(11)}$, R.~A.~Lacey$^{(1)}$, J.~Lauret$^{(1)}$,
C.~Law$^{(1)}$, M.A.~Lisa$^{(3)}$, H.~Liu$^{(4)}$, Y.M.~Liu$^{(14)}$,
R.~McGrath$^{(1)}$, Z.~Milosevich$^{(8)}$,
G.~Odyniec$^{(2)}$, D.L.~Olson$^{(2)}$,
S.~Y.~Panitkin$^{(4)}$, N.T.~Porile$^{(7)}$, G.~Rai$^{(2)}$, H.G.~Ritter$^{(2)}$,
 J.L.~Romero$^{(5)}$, R.~Scharenberg$^{(7)}$, L.~Schroeder$^{(2)}$,
B.~Srivastava$^{(7)}$, N.T.B~Stone$^{(2)}$, T.J.M.~Symons$^{(2)}$,
J.~Whitfield$^{(8)}$, T.~Wienold$^{(*2)}$, R.~Witt$^{(4)}$, L.~Wood$^{(5)}$,
and W.N.~Zhang$^{(14)}$
                        \\  (E895 Collaboration ) \\
P.~Danielewicz$^6$ and P.B.~Gossiaux$^{10}$ }

\address{
$^{(1)}$Depts. of Chemistry and Physics,
University \@ Stony Brook, New York 11794-3400 \\
$^{(2)}$Lawrence Berkeley National Laboratory,
Berkeley, California, 94720\\
$^{(3)}$Ohio State University, Columbus, Ohio 43210\\
$^{(4)}$Kent State University, Kent, Ohio 44242 \\
$^{(5)}$University of California, Davis, California, 95616 \\
$^{6}$ Michigan State University, East Lansing MI 48824-1321 \\
$^{(7)}$Purdue University, West Lafayette, Indiana, 47907-1396 \\
$^{(8)}$Carnegie Mellon University, Pittsburgh, Pennsylvania 15213\\
$^{(9)}$Brookhaven National Laboratory, Upton, New York 11973 \\
$^{10}$SUBATECH, Ecole des Mines, F-44070 Nantes, France  \\
$^{(11)}$University of Auckland, Auckland, New Zealand \\
$^{(12)}$Columbia University, New York, New York 10027 \\
$^{(13)}$St. Mary's College, Moraga, California  94575 \\
$^{(14)}$Harbin Institute of Technology, Harbin, 150001 P.~R. China \\
 }
\date{\today}

\maketitle
\newpage
\begin{abstract}
%\doublespacing

        We have measured the proton elliptic flow excitation function
for the Au + Au system spanning the beam energy range 2 -- 8 AGeV. The
excitation function shows  a transition from negative to positive
elliptic flow at a beam energy,   $E_{tr} \sim$ 4~AGeV.  Detailed
comparisons  with calculations from a relativistic Boltzmann-equation
are presented.  The comparisons suggest a softening  of the
nuclear equation of state (EOS) from a stiff form (K$\sim$~380 MeV) 
at low beam energies ($E_{Beam} \le 2$~AGeV)  to a softer form  
(K$\sim$~210 MeV) at higher energies ($E_{Beam}
\ge $ 4~AGeV ) where the calculated baryon density $ \rho \sim 4 \rho_0$.
\end{abstract}

%\narrowtext

        For many years, the investigation of the nuclear equation of
state~(EOS) has stood out as one of the primary driving forces  for heavy ion
reaction studies (e.g.~\cite{stocker86,qm96}).
Measurements of collective motion and, in particular, the elliptic flow
have been predicted to provide information crucial for
establishing the parameters of the EOS~\cite{oll92,dan95,sor97}.
%
%Measurements of collective motion and,
%in particular, the elliptic flow have been predicted to provide crucial
%information about the pressure build up in the early stages of mid-central
%heavy ion collisions~\cite{oll92,dan95,sor97}.
%
Theoretical conjectures have
also focused on the notion that a transition to the  quark-gluon plasma (QGP)
is associated with a ``softest point" in the EOS where the
pressure increase with temperature is much slower
than the energy density~\cite{hun95}.  Such
a softening of the EOS is predicted  to start  at quark-antiquark
densities comparable to those in the ground-state of nuclear
matter~\cite{lae96}, and also at relatively low temperatures if the baryon
density is driven significantly beyond its normal
value~$\rho_0$~\cite{dan98,bay76a}. At energies  of $1 \lesssim
E_{Beam} \lesssim 11$~AGeV, collision-zone matter
densities are expected up to $\rho \sim 6-8 \rho_0$\cite{dan98,boanli96}.
Such densities could very well result in conditions favorable to a
softening of the EOS. Therefore, it  is important
to investigate currently available elliptic flow data
[in  this energy range] to search for new insights into the
parameters of the EOS and for any indication of its softening.

Elliptic flow reflects the anisotropy of transverse particle emission at
midrapidity. For beam energies of 1--11~AGeV this anisotropy results from
a strong competition between ``squeeze-out" and
``in-plane flow"\cite{oll92,sor97,dan98}.
The magnitude and the sign of elliptic flow depend on two
factors:
(a)~the pressure built up in the compression stage compared to the energy
density, and (b)~the passage time of the projectile and target spectators.
The characteristic time for the development of expansion
perpendicular to the reaction plane can be estimated as $\sim R/c_s$, where
the speed of sound  $c_s =\sqrt{\partial p / \partial e}$, $R$~is the nuclear
radius, $p$~is the pressure and $e$ is the energy density. The passage time is
 $\sim 2R /(\gamma_0 \, v_0)$, where $v_0$ is the c.m. spectator velocity.
Thus the "squeeze-out" contribution should reflect the
ratio ${c_s / \gamma_0 \, v_0 } $~\cite{dan95} which is responsible
for the essentially logarithmic dependence of elliptic flow on the
beam energy for $\sim 1\le E_{beam} \le 11$ AGeV~\cite{dan98}.

	Recent calculations have made specific predictions
 for the beam energy dependence of elliptic flow for Au + Au collisions at
1--11 AGeV~\cite{dan98}. They  indicate a transition from
negative to positive elliptic flow at  a beam energy  $E_{tr}$, which  has
a marked sensitivity to the stiffness of the EOS.  In addition, they
suggest that a phase transition to the QGP should give a
characteristic signature in the elliptic flow excitation function due
to significant softening of the EOS.  In this Letter we present
an experimental elliptic flow excitation function for the Au + Au system
to establish $E_{tr}$ and to search for any hints of  a
softening of the EOS.

                The measurements were performed at  the Alternating Gradient
Synchrotron~(AGS) at the Brookhaven National Laboratory.  Beams of $^{197}$Au
 ($E_{Beam} = 2$, 4, 6, and 8~AGeV)\cite{beam_corr}
were used to bombard a $^{197}$Au target
of thickness calculated for a 3\% interaction probability.  Typical beam
intensities  resulted in $ \sim 10$ spills/min with $\sim 10^3$  particles per
 spill.  Charged reaction products were detected with the E895 experimental
setup which consists of  a time projection chamber (TPC)\cite{GRai90} and
a multisampling ionization chamber (MUSIC)\cite{Bauer97}.  The TPC  which was
located in the MPS magnet (typically at 1.0 Tesla)  provided good
acceptance and charge resolution for charged particles $-1<Z<6$ at all four
beam energies. However, unique mass resolution for $Z=1 $ particles
 was not achieved for all rigidities. The MUSIC device, positioned $ \sim$ 10
 m downstream of the TPC, provided unique charge resolution for fragments with
 $Z > 7$ for the 2 and 4~AGeV beams.  Data were taken with a trigger for
minimum bias and also for a bias toward central and mid-central collisions.
Results are presented here for protons measured in the TPC for mid-central
collisions.

         We use the second Fourier coefficient $v_2 = \langle \cos{2 \phi}
\rangle $,  to measure the elliptic flow or azimuthal asymmetry of the proton
distributions at midrapidity ($|y_{cm}| < 0.1$) ;
\begin{equation}
 {dN \over d\phi} \sim \left[ 1 + 2v_1\cos(\phi) +2v_2\cos(2\phi)  \right].
\label{Dist}
\end{equation}
Here, $\phi$ represents the azimuthal angle of an emitted proton relative to
the reaction plane.
The Fourier coefficient $\langle \cos{2 \phi} \rangle = 0$, $> 0$, and $< 0$
for zero, positive, and negative elliptic flow respectively. Measurements of
$v_1$ will be presented and discussed in a forthcoming paper\cite{hliu98}.

		Our analysis proceeds in two steps.  First, we determine the
reaction plane and  its associated dispersion for each beam energy. Second,
we generate  azimuthal distributions with respect to this experimentally
determined reaction plane and evaluate $\langle \cos{2 \phi} \rangle $. The
vector {\bf Q$_i$} = $\sum_{j\neq i}^{n}{w(y_j) \, {{\bf p}_j^t  / p_j^t }}$
is used to determine the azimuthal angle,  $ \Phi_{plane}$,  of the reaction
plane~\cite{dan85}. Here,  ${\bf p}_j^t$ and $y_j$ represent,  respectively,
the transverse momentum and the  rapidity of baryon~j (Z$\le 2$) in an event.
 The weight $ w (y_j)$ is assigned the  value ${ <p^x>\over <p^t>}$, where
$p^x$ is the transverse momentum in the reaction plane. $<p^x>$ is
obtained from the first pass of an iterative procedure.

	The dispersion of the reaction plane as well as biases associated with
 detector efficiencies plays a central role in flow
analyses\cite{dan87,oll97,Postkanzer98}. Consequently, in Fig.~\ref{fig2} we
show representative distributions for the experimentally determined
reaction-plane ($\Phi_{Plane}$), and the associated relative reaction-plane
distributions ($\Phi_{12}$). The distributions have been generated for
a mid-central impact parameter, i.e. multiplicities between 0.5
and 0.75 M$_{max}$. Here, M$_{max}$ is the
multiplicity corresponding to the point in the charged particle multiplicity
distribution where the height of the multiplicity distribution has fallen to
half its plateau value\cite{gut89}.  It is estimated that this multiplicity
range corresponds to an impact parameter range $\sim 5 - 7$~fm.
The $\Phi_{12}$ distributions (cf. Fig.~\ref{fig2}) which are important
for assessing the role of the reaction-plane dispersion, have been obtained via
 the subevent method~\cite{dan85}. That is, reaction planes
 were determined for two subevents constructed from each event; $\Phi_{12}$ is
 the absolute value of the relative azimuthal angle between these two
estimated reaction planes.  The essentially flat reaction plane distributions
shown in Fig.~\ref{fig2}a reflect rapidity and multiplicity-dependent
azimuthal efficiency corrections, applied  to take account of the
 detection inefficiencies of the TPC.  These corrections were obtained
by accumulating the laboratory azimuthal distribution of the particles (as a
function of rapidity and multiplicity) for all events and then including the
inverse of these distributions in the weights for the determination of the
reaction plane. The distributions shown in Fig.~\ref{fig2}a
confirm the absence of significant distortions which could influence
 the magnitude of the extracted elliptic flow.  The relative reaction-plane
distributions ($\Phi_{12}$) shown in Fig.~\ref{fig2}b indicate mean values
which increase with the beam energy from $<\Phi_{12}>/2~\sim~17.0^\circ$ at
2~AGeV to $\sim~36.1^\circ$ at 8~AGeV.  This increase suggests a
progressive deterioration in the resolution of the reaction plane with
increasing beam energy; however a reasonable resolution  is
maintained over  the entire energy range. The
$\Phi_{12}$ distributions serve as the basis for correcting the extracted
elliptic flow values as discussed below.

		In Fig.~\ref{fig3}, we show observed (or $\phi '$)
azimuthal distributions,
 for protons. The distributions, shown for several rapidity bins,
have been generated for the same mid-central impact
parameter range ($\sim 5 - 7$~fm) discussed  above. Several characteristic
features are exhibited
in Fig.~\ref{fig3}. For example, as one moves away from midrapidity, the
$\phi '$ distributions exhibit shapes commonly attributed to collective
sidewards flow. That is, for  $y > 0$, the distributions peak at $0^\circ$,
and,  for $y < 0$, they peak at  $\pm 180^\circ$. Fig.~\ref{fig3}  also shows
that these anisotropies decrease with increasing beam energy.

	The primary feature of the midrapidity distributions
contrasts with those  obtained at  other rapidities. At  2~AGeV,
two distinct peaks can be seen at  $-90^{\circ}$ and $+90^{\circ}$.  These
peaks indicate a clear signature for the ``squeeze-out" of nuclear matter
perpendicular to the reaction plane~\cite{dan87,gut89,dem90,swang96} or
negative elliptic flow. By contrast, at 6 and 8~AGeV, the midrapidity
distributions peak at  $0^{\circ}$, and $\pm180^{\circ}$.
This latter anisotropy pattern is expected  for positive elliptic
 flow. Thus, Fig.~\ref{fig3}c provides clear evidence for negative elliptic
flow at  2~AGeV, positive elliptic flow  for 6 and
8~AGeV, and near zero flow for $E_{Beam} = 4$~AGeV.

		In order to quantify the proton elliptic flow, it is necessary
 to suppress possible distortions arising from imperfect particle
identification (PId). It is relevant to reiterate here that unique
separation of $\pi^+$ and protons was not achieved for all rigidities.
To suppress such ambiguity we applied the following procedure.
First, we plot the
observed Fourier coefficient $\langle \cos{2 \phi } '\rangle $~vs.~$p_t$ with
$p_t$ thresholds which allow clean particle separation ($p_t \sim 1$~GeV/c).
We then extract the coefficients for the quadratic dependence of  $\langle
\cos{2 \phi } ' \rangle $ on $p_t$ (see inset in Fig.~\ref{fig4}).
These quadratic fits are restricted by the requirement
that $\langle \cos{2 \phi } ' \rangle = 0$
for  $p_t = 0$.  Second, we correct the proton $p_t$ distributions for possible
$\pi^+$ contamination by way of a probabilistic PId. The
latter probabilities were obtained by extrapolating the exponential tails
of the proton and $\pi^+$ rigidity distributions into the regions of overlap.
A weighted average (relative number of protons in a $p_t$
bin times the $\langle \cos{2 \phi } ' \rangle $ for that bin) was then
performed to obtain $\langle \cos{2 \phi } '\rangle $ for each beam energy.
Subsequent to this evaluation, we then use  the relative reaction plane
distribution at each beam energy (cf. Fig.~\ref{fig2}) to obtain dispersion
corrections for  the extracted  Fourier coefficients~\cite{dan85,oll97,dem90}.

		The relationship between the  $\langle \cos{2 \phi} '\rangle $
 (obtained with the estimated reaction plane) and the Fourier coefficient
$\langle \cos{2 \phi}\rangle$ relative to the true reaction plane is:
\begin{equation}
\langle \cos{2 \phi} '\rangle  = \langle \cos{2 \phi} \rangle
 \,  \langle \cos{2 \Delta \Phi} \rangle \, .
\label{Ft}
\end{equation}
where $\langle \cos{2 \Delta \Phi} \rangle$ is the correction
factor determined from the
$\langle \cos {\Phi_{12}} \rangle$~\cite{oll97}.
Following the prescription outlined in Ref.\cite{oll97}, we find
correction factors which range from  0.79
at 2~AGeV to  0.29 at 8~AGeV.  The  correction factors are summarized along
with ($\langle \cos {\Phi_{12}}\rangle$) in Table 1.

	The corrected elliptic flow values, $\langle \cos{2 \phi}\rangle$,
are represented by filled stars in Fig.~\ref{fig4}.  This excitation function
clearly shows an evolution from negative to positive elliptic flow within the
region  $2 \lesssim E_{Beam} \lesssim 8$~AGeV and points to an apparent
transition energy $E_{tr} \sim 4$~AGeV.  The solid and dashed curves
represent the results of model calculations described below. Since the value
of $E_{tr} $  is  predicted to be sensitive to the parameters of the
EOS\cite{dan98}, it is important to examine additional constraints on its
value. The inset in Fig.~\ref{fig4} shows the corrected
$\langle \cos{2 \phi } \rangle$ values as a function of~$p_t$ for protons.
The solid curves in the figure represent  quadratic fits to the
data (2 and 6 AGeV) which are  in agreement
with the predicted quadratic dependence of
$\langle \cos{2 \phi }\rangle$ on $p_t$~\cite{dan95,brill96}. 
Of greater significance is the fact that a comparison 
of the~$p_t$  dependence of the elliptic flow
for 2, 4, and 6 AGeV, provides further direct evidence that the sign of
elliptic flow changes as the beam energy is increased from 2 to 6~AGeV.
The essentially flat $p_t$ dependence shown for 4~AGeV is
consistent with $E_{tr} \sim 4$~AGeV.

        To interpret these data, extensive calculations have been
made to constrain the parameters of the EOS in the context of a newly
developed relativistic Boltzmann-equation model (BEM)~\cite{dan98,gos98}.
The phenomenological relativistic Landau theory of quasiparticles~\cite{bay76}
 serves as a basis for the model which has nucleon, pion,  $\Delta$ and $N^*$
resonance degrees of freedom as well as momentum dependent forces.
Calculations were performed for both a soft ($K=210$~MeV), and
a stiff ($K=380$~MeV) EOS for the same rapidity and impact parameter
selections applied to the data.

        The elliptic flow excitation functions (calculated for free protons)
are compared to the experimental data in Fig.~\ref{fig4}. The dashed and
solid curves represent the results for a stiff and a soft EOS respectively.
In addition to the data from the present
experiment (filled stars),  Fig.~\ref{fig4} also
shows experimental results for Au~+~Au reactions at 1.15 A~GeV\cite{eosx}
(filled triangle) and 10.8~A~GeV\cite{e877} (filled circle). The experimental
data are compatible with the excitation function predicted for a stiff
EOS at beam energies $1 \lesssim E_{Beam} \le 2$~AGeV.  By contrast,
the data show good agreement with the predictions for a soft EOS
for $4\le E_{Beam} \lesssim 11$~AGeV.
This pattern is consistent with a softening of the EOS in semicentral
collisions of Au~+~Au  at $\sim$~4~AGeV. The calculated densities at
maximum compression for these energies are of the
order of $\sim 4 \, \rho_0$ for the stiff EOS.

        In summary, we have measured an elliptic flow excitation function for
mid-central collisions of Au + Au  at 2, 4, 6, and 8~AGeV. The excitation
function exhibits a transition from negative to positive elliptic flow  with
$E_{tr} \sim 4$~AGeV.   Detailed comparisons of these
elliptic flow data  have been made with calculated results from a relativistic
 Boltzmann-equation calculation. Within the context of a simple parametrization
 of the EOS, the calculations suggest an evolution from a stiff EOS
(K$\sim 380$ MeV) at low beam energies ($\le 2$ AGeV) to a softer  EOS
(K$\sim 210$ MeV) at higher beam energies ($4 \le E_{Beam} \lesssim 11$~AGeV).
 Such a softening of the EOS could result from a number of effects, the most
intriguing of which is the possible onset of a nuclear
phase change\cite{dan98}. On the other hand, it should be noted that
transport models have failed to reproduce low energy
"squeeze-out" data with a single incompressibility
constant\cite{swang96}.
Thus, additional experimental signatures as well as calculations
based on other models will be necessary to test the detailed
implications of these results. Nevertheless, the
results presented here, clearly show that elliptic flow
measurements can provide an important constraint on the
EOS of high density nuclear matter.

%\acknowledgements

        This work was supported in part by the U.S.\ Department of
Energy under grants DE-FG02-87ER40331.A008, DE-FG02-89ER40531,
DE-FG02-88ER40408, DE-FG02-87ER40324, and contract DE-AC03-76SF00098; by the
US National Science Foundation under Grants
No.\ PHY-98-04672, PHY-9722653, PHY-96-05207,
PHY-9601271, and PHY-9225096; and by the University of Auckland Research
Committee, NZ/USA Cooperative Science Programme CSP 95/33.
\newline
$^*$ Feodor Lynen Fellow of the Alexander v. Humboldt Foundation.
%\newpage

\newpage

\begin{figure}[h]
\centerline{\epsfysize=1.2in \epsffile{tables.epsi}}
\vspace*{.4in}
%\caption{ Dispersion correction factors for each beam energy.}
\end{figure}
%
%\begin{figure}
%\centerline{\epsfysize=4.0in \epsffile{fig1.eps}}
%\vspace*{.4in}
%\caption{
%                Schematic illustration of the collision of
%two Au nuclei at relativistic energies.   Time shots are shown for an~instant
%before the collision~(a), early in the collision~(b), and late in the
%collision~(c).}
%\label{fig1}
%\end{figure}
%

\newpage

\begin{figure}
%\centerline{\epsfxsize=4.5in \epsffile{fig2.ps}}
\centerline{\epsfxsize=3.5in \epsffile{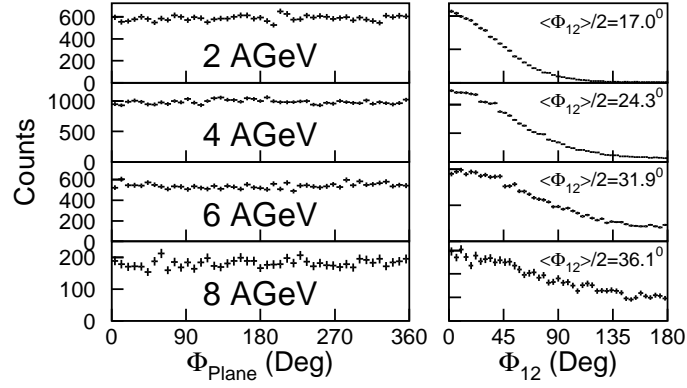}}
\vspace*{.4in}
\caption{ Experimentally determined (a) reaction-plane
($\Phi_{Plane}$) distributions, and (b) the associated relative reaction-plane
 distributions ($\Phi_{12}$) for 2, 4, 6, and 8 AGeV Au~+~Au.
The reaction plane distributions include efficiency corrections
for the TPC (see text).
}
\label{fig2}
\end{figure}

\newpage

\begin{figure}
\centerline{\epsfxsize=3.5in \epsffile{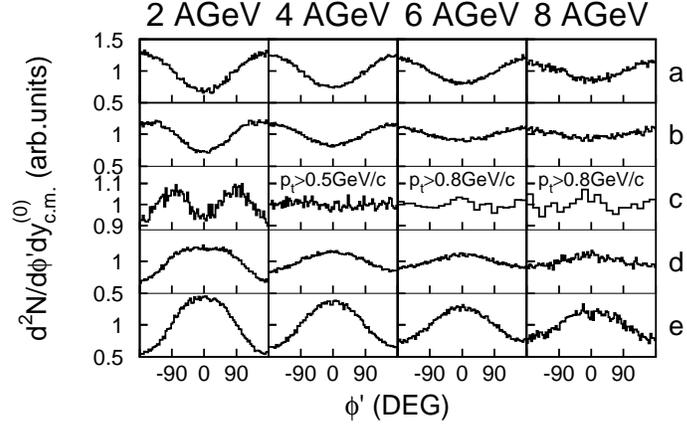}}
\vspace*{.4in}
\caption{ Azimuthal distributions (with respect to the reconstructed
reaction plane) for 2, 4, 6, and 8 AGeV Au + Au. Distributions are shown for
(a) $-0.7<y_{cm}<-0.5$, (b) $-0.5<y_{cm}<-0.3$, (c) $-0.1<y_{cm}<0.1$,
 (d) $0.3<y_{cm}<0.5$, and (e) $0.5<y_{cm}<0.7$.
The mid-rapidiy selections for 4 - 8 AGeV  also include a
transverse momentum selection as indicated.}
\label{fig3}
\end{figure}

\newpage
\newpage

\begin{figure}
\centerline{\epsfxsize=3.5in \epsffile{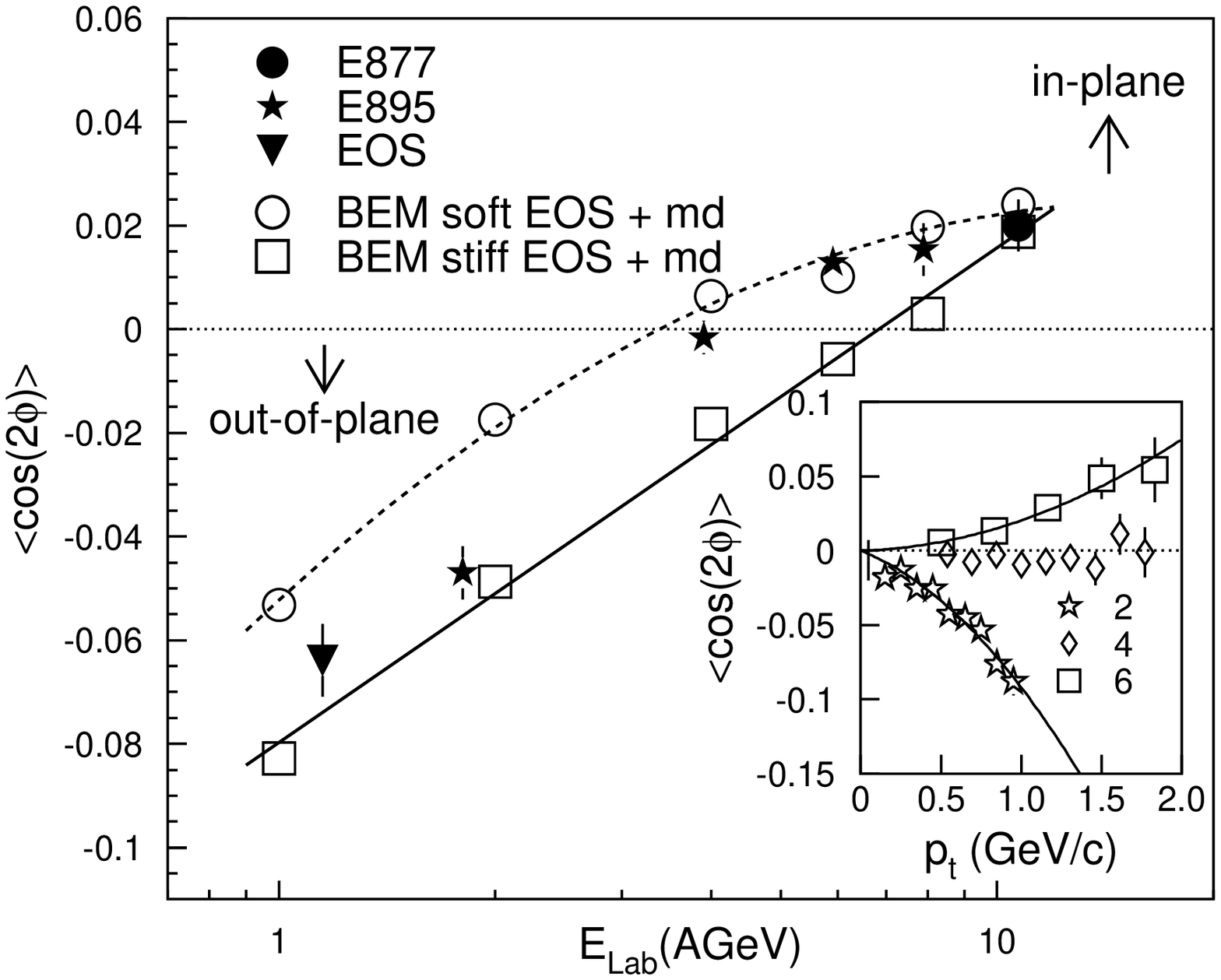}}
\vspace*{.4in}
\caption{ Elliptic flow excitation function for Au + Au. The filled symbols
 represent the experimental data as indicated.  The dashed curve
(open circles) and the solid curve (open squares) represent the
calculated excitation functions for a soft and
a stiff EOS (both with momentum dependence) respectively.
The inset shows the [dispersion corrected] transverse momentum dependence of
the elliptic flow for the 2, 4 and 6 AGeV beams.
}
\label{fig4}
\end{figure}

\begin{references}
\bibitem{beam_corr} After correction for energy loss in the beam line
		    and the target, the respective energies are 1.8,
		    3.9, 5.9, and 7.9 AGeV.
\end{references}
\end{document}